\documentclass{llncs}

\usepackage{amssymb}
\usepackage{amsmath}
\usepackage{epic}
\usepackage{eepic}
\usepackage{graphicx}
\DeclareSymbolFontAlphabet{\mathbb}{AMSb}
\usepackage{Vmargin}
\setmargins{1in}{1in}{6.5in}{9in}{0pt}{0pt}{12pt}{42pt}
\usepackage{times}
\usepackage{amsfonts}
\usepackage{eucal}

\begin{document}

\title{For-profit mediators in sponsored search advertising}

\author{ Sudhir Kumar Singh\inst{1,}\thanks{This work was done while the author was working for Ilial Inc.. The financial
support from Ilial Inc. is highly acknowledged.} \and  Vwani P. Roychowdhury\inst{1,2}
 \and Himawan Gunadhi\inst{2} \and Behnam A. Rezaei\inst{1,2}}
\institute{ Department of Electrical Engineering, University of California, Los Angeles, CA 90095.\\
\and Ilial Inc., 11943 Montana Ave. Suite 200, Los Angeles, CA 90049. \\
{\it \{sudhir,vwani,gunadhi,behnam\} @ ilial.com}}

\date{\today}
\maketitle

\begin{abstract}

A mediator is a well-known construct in game theory, and is an entity that plays on behalf
of some of the agents who choose to use its services, while the rest of the agents
participate in the game directly.  We initiate a game theoretic study of sponsored search
auctions, such as those used by Google and Yahoo!, involving {\em incentive driven}
mediators. We refer to such mediators as {\em for-profit} mediators, so as to distinguish
them from  mediators introduced in prior work, who have no monetary incentives, and are
driven by the altruistic goal of implementing certain desired outcomes. We show that in
our model, (i) players/advertisers can improve their payoffs by choosing to use the
services of the mediator, compared to directly participating in the auction; (ii) the
mediator can obtain monetary benefit by managing the advertising burden of its group of
advertisers; and (iii) the payoffs of the mediator and the advertisers it plays for are
compatible with the incentive constraints from the advertisers who do dot use its services. A simple intuition
behind the above result comes from the observation that the mediator has more information
about and more control over the bid profile than any individual advertiser, allowing her to reduce the payments
made to the auctioneer, while still maintaining incentive constraints. 
Further, our results indicate that there are significant opportunities for diversification
in the internet economy and we should expect it
to continue to develop richer structure, with room for different types of
agents to coexist. 

\end{abstract}

\section{Introduction}

With the growing popularity of the Web for obtaining information via search,
\emph{sponsored search} advertising, where advertisers pay to appear alongside the
algorithmic results, has become a significant business model and is responsible for the
success of internet giants such as Google and Yahoo!   The statistics show that the growth
of the overall online advertising market has been around 30\% every year, as compared to
the 1-2\% of the traditional media. The first quarter of 2007 also saw a tremendous
increase in revenue from online advertising that is 26\% over that in 2006. Search remains
the largest revenue format, accounting for more than  40\% of the 2006 full year revenues
of around \$17 billion.

In a search-based advertising format, the Search Engine  allocates the available
advertising space using an auction, where individual advertisers bid upon specific
keywords. When a user queries for a keyword, the search engine (the auctioneer) allocates
the advertisement space to the bidding merchants based on their bid values and their
estimated fitness values. Usually, the ads appear in a separate section of the page
designated as ``sponsored search results,'' which is located  above or to the right of the
organic/algorithmic results. Each position in such a list of sponsored links is called a
{\it slot}.  Generally, users are more likely to notice and click on a higher ranked slot,
leading to more traffic for the corresponding advertisers. Therefore, advertisers prefer
to be in higher ranked slots and compete for them. In a popular scheme, known as the
Cost-Per-Click (CPC) or the Pay-Per-Click (PPC) model, whenever a user clicks on an ad,
the corresponding advertiser pays an amount specified by the auctioneer.

From the above description, we can note that after merchants have bid for a specific
keyword, when that keyword is queried, auctioneer follows two steps. First, she allocates
the slots to the advertisers. Normally, this allocation is done using some {\it ranking
function}. Secondly, she decides, through some {\it pricing scheme}, how much a merchant
should be charged if the user clicks on her ad and in general this depends on which slot
she was assigned, on her bid and that of others.
In the auction formats for sponsored search, there
are two ranking functions, namely {\it rank by bid} (RBB) and  {\it rank by revenue}(RBR)
and there are two pricing schemes, namely {\it generalized first pricing}(GFP) and  {\it
generalized second pricing}(GSP) which have been used widely. In RBB, bidders are ranked
solely according to their bid values. The advertiser with the highest bid gets first slot,
that with the second highest bid gets the second slot and so on. In RBR, the bidders are
ranked according to the product of their bid value and {\it quality score}. The quality
score represents the merchant's relevance to the specific keyword, which can basically be
interpreted as the possibility that her ad will be viewed if given a slot irrespective of
what slot position she is given. In GFP, the bidders are essentially charged the amount
they bid and in GSP they are charged an amount which is enough to ensure their current
slot position. For example, under RBB allocation, GSP charges a bidder an amount  equal to
the bid value of the bidder just below her.

Formal analysis of such sponsored search advertising models has been done extensively in
recent years, from algorithmic as well as from game theoretic
perspectives\cite{EOS05,MSVV05,Lah06,AGM06,Var06,LP07,MNS07}. For example, the existence
of different types of incentive-driven Nash equilibria has been established. Further, the
notion of a {\it mediator} in such position auctions has also been discussed\cite{med}. A
mediator is a reliable entity, which can play on the behalf of agents in a given game,
however it can not enforce the use of its services, and each agent is free to participate
in the game directly. In the paper by Ashlagi et al.\cite{med} and the references therein,
the motivation for the use of mediator comes from the search of means to implement
particular outcomes, such as VCG, in a given mechanism such as RBR with GSP. However, the
mediators considered so far are altruistic in nature and have no incentives, and in
particular, their only goal is to implement certain outcomes despite the financial cost
incurred. As we know, the marketplace is mostly about incentives- a game between selfish
agents- and it would be interesting to study mediators which are not altruistic.

In our present work, we initiate a study of mediators in sponsored search auctions, which
may not be altruistic in nature. We call such mediators as {\it for-profit} mediators and
show that advertisers can improve their payoffs by using the services of the mediator
compared to directly participating in the auction and mediator can also obtain monetary
benefit by managing the advertising burden of its advertisers and in fact at the same time
being compatible with incentive constraints from the advertisers who do dot use its service. 
The simple
intuition behind the above result comes from the observation that since the mediator has
more information about and more control over the bid profile
than any individual advertiser, she could possibly
modify their bids, before reporting to the auctioneer (search engine), in a manner to
improve their payoff and could retain a fraction of the improved payoff. Thus, our results
show that mediators can play a significant role in sponsored search auctions, and can
potentially impact the revenues earned by the auctioneer.

\section{Definitions and Model Setup}
In a formal setup, there are $K$ slots to be allocated among  $N$ ($\geq K$) bidders. A
bidder $i$ has a true valuation $t_i$ (known only to the bidder $i$) for the specific
keyword and she bids $v_i$. The expected {\it click through rate} (CTR) of an ad put by
bidder $i$ when allocated slot $j$ has the form $\gamma_j e_i$ i.e. separable in to a
position effect and an advertiser effect. $\gamma_j$'s can be interpreted as the
probability that an ad will be noticed when put in slot $j$ and it is assumed that
$\gamma_1 > \gamma_2
>\dots > \gamma_K > 0$. $e_i$ can be interpreted as the probability that an ad put by
bidder $i$ will be clicked on if noticed and is refered as the {\it relevance} of bidder
$i$. This is the quality score used in the RBR allocation rule mentioned earlier. The
payoff/utility of bidder $i$ when given slot $j$ at a price of $p$ is given by
$e_i\gamma_j (t_i - p)$ and they are assumed to be rational agents trying to maximize
their payoffs. As of now, Google as well as Yahoo! uses schemes closely modeled as RBR
with GSP. The bidders are ranked according to $e_iv_i$ and the slots are allocated as per
this ranks. For simplicity of notation, assume that the $i$th bidder is the one allocated
slot $i$ according to this ranking rule, then $i$ is charged an amount equal to
$\frac{e_{i+1} v_{i+1}}{e_i}$.

Let the bid profile without any mediation be (under RBB) $v_1 \geq v_2 \geq \dots \geq v_L
> v_{L+1} > v_{L+2} > \dots > v_K > v_{K+1}$. Now suppose that the bidders $1,2,\dots,L$
manage their bidding process through a same third party i.e. the mediator. Since the
mediator has much more information about and more control over the bid profile 
than the individual advertisers it
is likely that she could 
modify the associated bids in a
manner to increase her payoff and that of the associated advertisers. For example, the
mediator can simply bid $v_1^{'}=v_1, v_2^{'}=v_3^{'}= \dots = v_L^{'}=v_L$ and she pays
(under GSP) an amount $(L-1)v_L + v_{L+1}$  which is much smaller than $\sum_{j=1}^L
v_{j+1}$ without mediation. Mediator can distribute a part of this profit to the
associated advertisers and therefore those advertisers along with the mediator profit at
the expense of revenue loss by the auctioneer. This is the basic intuition behind why
advertising via mediation can be good to the advertisers and the advertisers can stick to
their traditional media companies even for advertising in sponsored search and other such
auctions. However, the above intuition is ofcourse not a formal game-theoretic argument
why the collusion via mediation will work as we do also need to argue that the other
advertisers (the advertisers who do not advertise via the mediator) still do not have
incentives to change their slot positions.  In the following we present a game theoretic
analysis for the position auctions via mediation and show that the intuition given above
is indeed true.

We consider the case where there is only one mediator and the analysis in
the other cases essentially remains similar. The advertisers who bid via the mediator will
be called {\it $M$-bidders} and all other advertisers will be called {\it $I$-bidders}.
The essential features of the position auctions via mediation is:

\begin{itemize}

\item $M$-bidders report their bids to the mediator.

\item $M$-bidders do not want to change the positions they get via directly reporting to the auctioneer\footnote{Relaxing this condition gives more freedom to the mediator and she could possibly do even better by changing their positions as illustrated later in the paper, however advertisers
might not like to go down in slot position due to decrease in traffic as well as branding impression value.} however they give mediator
the right to change their bids before reporting to the auctioneer for a potential increase in their payoffs.

\item Mediator chooses a suitable set of bids for the associated advertisers and report  accordingly to the auctioneer
      on behalf of them.

\item $I$-bidders report their bids to the auctioneer directly.

\end{itemize}

\section{Designing for-profit mediators}

Let us first consider the RBR (rank by revenue) scheme with GSP(generalized second price)
currently being used by Google and Yahoo!. The advertisers are ranked according to $r_i =
e_i v_i$ where $e_i$ is the relevance (quality score). Let us name the advertisers by this
ranking i.e. $r_1 > r_2 > \dots > r_L > r_{L+1}> \dots > r_K >r_K > \dots > r_N$,
therefore the $i$th bidders pays $\frac{e_{i+1}v_{i+1}}{e_i} = \frac{r_{i+1}}{e_i}$ under
GSP.
Let us first analyze the incentive and revenue properties for the case where the top $L$ advertisers are the $M$-bidders.
We will be interested in a Walrasian type of equilibria of the associated game, which is called {\it symmetric} Nash equilibria(SNE)
as proposed by Varian\cite{Var06} and Edelman et al\cite{EOS05}. However, similar analysis can be
done for non-symmetric Nash equilibria and as we will note later,
the mediator and the advertisers might be even more better off
in the case of non-symmetric Nash equlibria.
Under SNE, the bidders have no incentive to change to another
positions even at the current price paid by the bidders currently at that position. Note that this is a stronger condition
than usual Nash equilibrium condition which for the case of moving to higher position
requires the defecting bidder to pay the bid of the advertiser
holding the position currently, which is more than the price paid by her under GSP.
The bids $v_1, v_2, \dots, v_L,v_{L+1}, \dots v_K \dots $ are at the SNE of the auction without any mediation,
therefore in the original game the bidders have no incentives to defect from their current positions\footnote{It is reasonable to assume this as
the auction process has been going for a while now. Further, this requirement can be relaxed- the mediator first 
bids on the behalf of the M-bidders to figure out and evolve to an equilibrium before implementing
her strategy to modify their bids.}. 
Now the problem is to how should the mediator modify the bids of $M$-bidders so as to maintain the
same incentives for the $I$-bidders and to improve her and M-bidders' payoffs. Here, mediator's
payoff is defined to be a fixed fraction of the total improvement in payoffs from the M-bidders 
over what they could have obtained without using her service, up to an additive constant.  

Let the mediator modify the bids as $r_i^{'}=r$ for $i=1,2,\dots,l$ and $r_i^{'}=r_i$ for $i=l+1, \dots, L$, then what $r$ and $l$
should she choose\footnote{The bids will actually be modified so that $r_i^{'}=r + (L-i) \epsilon$ for an infinitesimally small $\epsilon > 0$.
In practice, this $\epsilon$ can not be less than $\$0.01$, however for the purpose of analysis, as in earlier works, 
we assume that it is a continuous parameter that 
can be made infinitesimally small.}. The auctioneer sees the bid profile
$r \geq r \geq \dots \geq r > r_{l+1} > \dots >r_L > r_{L+1} > \dots > r_K > \dots $.
Since the original bid profile was at SNE, no $j \geq L+1$ would like to deviate to any other position $s$ for $l \leq s \leq K+1$.
Now, only position a $ j \geq L+1$ can deviate to is the position $1$ (she can not do better than this by moving to $2, \dots, l-1$
for she will be paying same price to get less clicks).

The condition that the $I$-bidders do not want to move to position $1$ is

\begin{displaymath}
\begin{array}{ll}
\gamma_j e_j(t_j - \frac{r_{j+1}}{e_j}) \geq \gamma_1 e_j (t_j - \frac{r}{e_j}) &  \forall j \geq L+1 \\
\gamma_j (e_jt_j - r_{j+1}) \geq \gamma_1 (e_j t_j - r) &  \forall j \geq L+1 \\
\therefore  r \geq (1 - \frac{\gamma_j}{\gamma_1}) e_j t_j + \frac{\gamma_j}{\gamma_1} r_{j+1} &  \forall j \geq L+1
\end{array}
\end{displaymath}

 Let
\begin{displaymath}
 r^{*} = \max_{j \geq L+1} \{(1 - \frac{\gamma_j}{\gamma_1}) e_j t_j + \frac{\gamma_j}{\gamma_1} r_{j+1}\}
\end{displaymath}
then any selection of $r$ such that $ r \geq r^{*}$ and $r> r_{l+1}$ is fine at SNE and the mediator chooses
$r=r^{*}$ and an $l$ such that $ r_l \geq r > r_{l+1}$. Note that such an $l \geq 2$ always exists as $ r \leq r_2$
(for the $I$-bidders did not want to move to position $1$ in the original game at SNE)\footnote{Of course, the mediator will not be able
to choose such a $r$ and $l$ all at once and will rather evolve to it by trying suitable $r_l$'s.},\footnote{Sometimes for example when
$l=2$ in the above, the mediator could possibly do even better by modifying bids as
$ r_1 > r_2 > r \geq r \geq \dots \geq r > r_{l+1} > \dots >r_L > r_{L+1} > \dots > r_K > \dots $ and so on.}.

The mediator now pays $(\sum_{j=1}^{l-1} \gamma_j )r + \sum_{j=l}^{L} \gamma_j r_{j+1}$ on behalf of the $M$-bidders and the
net total gain for mediator is
\begin{displaymath}
\sum_{j=1}^{l-1} ( r_{j+1} - r) \gamma_j
\end{displaymath}
 and the mediator can distribute a fraction of this to the associated advertisers.
It is clear that this is at the expense of the loss in the revenue of the auctioneer.
Note that in the case when there are only $M$-bidders and no $I$-bidders, the
auctioneer gets the minimum price set for all the slots. Now let us illustrate
the above analysis by an example listed in Table \ref{med1} wherein
the bid profile $\{r_i\}$ is first verified to be at symmetric Nash equlibrium in Table \ref{med2} by recalling that
to verify this we need only check the equilibrium condition for one slot up and one slot down positions
(locally envy free property)  and finally a suitable $r$ is chosen in Table \ref{med3}.

\begin{center}
\begin{table*}
\hspace{2in}
\begin{tabular}{|c|c|c|c|c|c|c|c|c|c|}
\hline
$i$  & 1 & 2 & 3 & 4 & 5 & 6 & 7 & 8 & 9  \\
\hline
$\gamma_i$ & 1 & 0.6 & 0.5 & 0.4 & 0.3 & 0.2 & 0.15 & 0.10 & 0 \\
\hline
$e_i t_i$ & 26 & 22 & 20 & 18 & 17 & 15 & 12 & 12 & 9 \\
\hline
$r_i=e_iv_i$ & 25 & 20 & 16 & 15 & 14 & 13 & 11 & 10 & 9 \\
\hline
$e_iPPC_i$ &  20 & 16 & 15 & 14 & 13 & 11 & 10 & 9 & 0 \\
\hline
$r_i^{'}=e_i v_i^{'}$ & 14.2 & 14.2 & 14.2 & 14.2 & 14 & & & &  \\
\hline
reduced $e_iPPC_i$ & 14.2 & 14.2 & 14.2 & 14 & 13 & & & &\\
\hline
\end{tabular}
\caption{Position based CTRs, true valuations, bid profile, and modified bid profile when $\{1,2,3,4,5\}$ are the $M$-bidders ($PPC_i$ denotes payment
per click by the bidder $i$).}
\label{med1}
\end{table*}
\end{center}

\begin{center}
\begin{table*}
\begin{tabular}{|c|c|c|c|c|}
\hline
position $j$ & payoff: & payoff by defecting to $j-1$:  & payoff by defecting to $j+1$: & SNE condition satisfied \\
&  $u_j=\gamma_j (e_jt_j -r_{j+1})$ &  $u_j^{j-1}=\gamma_{j-1} (e_jt_j -r_j)$ &  $u_j^{j+1}=\gamma_{j+1} (e_jt_j -r_{j+2})$ &  (YES/NO) \\
\hline
1 & 1 (26 -20)= & & 0.6 (26-16) = & \\
  &  6          & & 6 & YES \\
\hline
2 & 0.6 (22 -16)= & 1(22-20)= & 0.5 (22-15) = & \\
  &  3.6          & 2         & 3.5  & YES \\
\hline
3 & 0.5 (20 -15)= & 0.6 (20-16)= & 0.4 (20-14) = & \\
  &  2.5         & 2.4        & 2.4  & YES \\
\hline
4 & 0.4 (18 -14)= & 0.5 (18 -15)= & 0.3 (18-13)=& \\
  &  1.6         & 1.5 & 1.5 & YES \\
\hline
5 & 0.3 (17-13)= & 0.4 (17-14) = & 0.2(17-11)= & \\
  &  1.2         & 1.2 & 1.2 & YES \\
\hline
6 & 0.2 (15-11)= & 0.3 (15-13)= & 0.15(15-10)= & \\
  & 0.8          & 0.6          & 0.75 & YES \\
\hline
7 & 0.15 (12-10)= & 0.2 (12-11)= & 0.10 (12-9)= & \\
  & 0.3           & 0.2          & 0.3 & YES\\
\hline
8 & 0.10 (12-9)= & 0.15 (12-10)= & 0 (12-0) = & \\
  &   0.3        & 0.3           & 0 & YES \\
\hline
9 & 0            & 0.10 (9-9)= 0 & 0 & YES \\
\hline
\end{tabular}
\caption{Verifying the SNE conditions}
\label{med2}
\end{table*}
\end{center}

\begin{center}
\begin{table*}
\hspace{2in}
\begin{tabular}{|ccc|c|c|c|c|c|}
\hline
& $j$ & & $s_j$ &  $r^{*}$ & $r$ & $l$ & improved payoff: \\
&  & & $=e_jt_j - u_j$ &   & &  &$\sum_{j=1}^{l-1} ( r_{j+1} - r) \gamma_j$ \\
\hline
& 6 &  & 15-0.8 = & & & & \\
& &    &  14.2    & & & &\\

& 7 &   & 12-0.3=  & 14.2 & 14.2 & 4 &  7.28 \\
& &    & 11.7     &      &  &   &  \\

& 8 &   & 12-0.3 = & & & & \\
& &    &  11.7   & & & & \\

& 9  &  & 9-0 = & & &  &\\
  & &  &  9    & & & & \\
\hline
\end{tabular}
\caption{Computing $r$ and $l$: $s_j:=(1 - \frac{\gamma_j}{\gamma_1}) e_j t_j + \frac{\gamma_j}{\gamma_1} r_{j+1}=
e_jt_j-\gamma_j(e_jt_j-r_{j+1})=e_jt_j-u_j$ (as $\gamma_1 =1)$ }
\label{med3}
\end{table*}
\end{center}

If the bid profile is at non-symmetric Nash equilibrium then the mediator might do better
as she can modify the bids as
 $r_i^{'}=r$ for $i=2,\dots,l$ and $r_i^{'}=r_i$ for $i=1, l+1, \dots, L$ and the condition on $r$ now is
\begin{displaymath}
 r \geq \max_{j \geq L+1} \{(1 - \frac{\gamma_j}{\gamma_2}) e_j t_j + \frac{\gamma_j}{\gamma_2} r_{j+1}\}.
\end{displaymath}

Now let us consider the case when the $M$-bidders are not necessarily the top ones but
$ l+1, l+2, \dots, l+L$. The mediator modifies the bids as
$r_j^{'}=r$ for $j=l+2, \dots, l+s-1$ for some $s \leq L$ and $r_j^{'}=r_j$ otherwise, therefore
the auctioneer sees the bid profile
$r_1 > r_2 > \dots > r_l > r_{l+1} > r \geq r \dots \geq r > r_{l+s} > \dots > r_{l+L} > r_{l+L+1} > \dots $.

As in the earlier analysis, at SNE, the only condition that need to be checked is that no $j \leq l$ or $j \geq l+L+1$
would want to deviate to the $(l+1)$th position. Therefore, we must have for  $j \leq l$ and  $j \geq l+L+1$ ,
\begin{displaymath}
\begin{array}{l}
\gamma_j e_j(t_j - \frac{r_{j+1}}{e_j}) \geq \gamma_{l+1} e_j (t_j - \frac{r}{e_j})  \\
\gamma_j (e_jt_j - r_{j+1}) \geq \gamma_{l+1} (e_j t_j - r) \\
\therefore  r \geq (1 - \frac{\gamma_j}{\gamma_{l+1}}) e_j t_j + \frac{\gamma_j}{\gamma_{l+1}} r_{j+1} .
\end{array}
\end{displaymath}
 Let
\begin{displaymath}
 r^{*} = \max_{\{j \leq l\} \cup \{ j \geq l+L+1 \}} \{(1 - \frac{\gamma_j}{\gamma_{l+1}}) e_j t_j + \frac{\gamma_j}{\gamma_{l+1}} r_{j+1}\}
\end{displaymath}
then clearly $r^{*} \leq r_{l+2}$ and choosing any $ r \geq r^{*}$ and an $s$ such that $r_{l+s-1} \geq r > r_{l+s}$ is fine at SNE.
However, in this case or in the case when $M$-bidders are the top ones, such a $r$ to improve their payoffs might not always exist
as can been seen by considering the example mentioned above when $M$-bidders are $\{2,3,4,5\}$.
However, mediator could possibly improve even in these cases, if the $M$-bidders do not mind moving up in positions,
as she does not neccesarily have to satisfy the incentive constraints from higher position $I$-bidders to not change their current 
positions.  

Similar analysis holds when there are different groups of $M$-bidders such that all advertisers in a group bid
via the same mediator whereas different groups may have different mediators associated with them.
In fact, in this case the maximization will be over smaller sets and the mediators could possibly do even better.

A possibility not analyzed above is that whether the mediator can do better by moving the positions of
the advertisers either individually or sliding them all together. Consider the example given earlier and
let mediator slide every $M$-bidder one slot down by modifying the bid profile so that
$r_i^{'}=12$ for all M-bidders as shown in the Table \ref{med4}. It can be verified as
before that it is still at SNE and in fact in this case mediator's payment on behalf of M-bidders is much lesser
compared to the earlier case. This suggests that indeed the mediator could do better by
moving slot positions. However, advertisers might not like to change positions, at least not to the lower slots due to
associated branding impression values coming from higher slots and even though their payoff might increase by allowing so,
they might not like to lose in terms of traffic which decreases by going down.

\begin{center}
\begin{table*}
\hspace{2in}
\begin{tabular}{|c|c|ccccc|c|c|c|}
\hline
position $j$  & 1 & 2 & 3 & 4 & 5 & 6 & 7 & 8 & 9 \\
\hline
bidder $i$ assinged to $j$ & 6 &  1 & 2 & 3 & 4 & 5 & 7 & 8 & 9 \\
\hline
$e_i t_i$ & 15 & 26 & 22 & 20 & 18 & 17  & 12 & 12 & 9 \\
\hline
$r_i=e_iv_i$ &13 &  25 & 20 & 16 & 15 & 14 & 11 & 10 & 9 \\
\hline
$e_i PPC_i$ & 11 & 20 & 16 & 15 & 14 & 13 & 10 & 9 & 0 \\
\hline
$r_i^{'}=e_i v_i^{'}$ &  & 12 & 12 & 12 & 12 &12 & & &  \\
\hline
reduced $e_i PPC_i$ &  & 12 & 12 & 12 & 12 & 11& & & \\
\hline
improved payoff: & & & & 22.8 & & & & & \\
\hline
\end{tabular}
\caption{Sliding positions could improve $M$-bidders' and mediator's payoffs}
\label{med4}
\end{table*}
\end{center}

For-profit mediators for other mechanisms can also be designed.
In particular, we discuss for-profit mediators for truthful mechanisms in the following. 
Truthful mechanisms are considered to be very desirable from the advertisers' perspective
since truth-telling is a dominant strategy for every one and the advertises do not need
to be sophisticated to play the auction game. However, as we argue below it is 
more vulnerable to for-profit mediation and even the mediators
need not be sophisticated in this case, unlike the ones discussed earlier in this paper.
In regard to position auctions, Aggarwal et al.\cite{AGM06} presented a truthful 
mechanism called {\it laddered auction},
which is compatible with a given weighted ranking function such as RBR, and is the unique truthful auction
given this ranking function. 
Now, if the mediator modifies the bids of the $M$-bidders in a manner so that their slot 
positions (i.e. ranks) do not change, the $I$-bidders still report truthfully as its 
a dominant strategy. The mediator could choose such a minimum possible 
bid profile to get the best improvement in payoffs and in particular modifying 
every $M$-bidder's bid to a value only infinitesimally more than just enough to retain the position of the least ranked $M$-bidder
suffices.

\section{Concluding Remarks and Future Work}

Sponsored search advertising is a significant growth market and is witnessing rapid growth and
evolution. The analysis of the underlying models has so far primarily focused on the
scenario, where advertisers/bidders interact directly with the auctioneers, i.e., the
Search Engines and publishers. However, the market is already witnessing the spontaneous
emergence of several categories of companies who are trying to mediate or facilitate the
auction process. For example, a number of different AdNetworks have started proliferating,
and so have companies who specialize in reselling ad inventories. Hence, there is a need
for analyzing the impact of such incentive driven and for-profit agents, especially as
they become more sophisticated in playing the game.

Our results show that there are significant opportunities for diversification in the
market and the emergence of incentive-driven equilibria. Thus, we should expect the
adword auction market to continue to develop richer structure, with room for different types of
agents to coexist. Another implication of our results applies to the
traditional media. Publishers of traditional media, such as newspapers and network TV and
radio, have seen significant declines in their audience market shares, as more people have
shifted to the Web as the source for information and entertainment. Their advertising
revenues have decreased significantly as well. Our results show that one way these
traditional media players can retain the loyalty of their advertisers is to manage their
online auctions! By mediating their auctions, they can provide better payoffs to their
clients, and thus prevent them from switching allegiance to the online giants, such as
Google and Yahoo! Other than creating a new revenue source for the traditional media
businesses, it would allow them to retain their own networks and give them precious time
to reposition themselves and figure out the best possible ways to take their content
online, and compete effectively in a new market space.

Our present work on the diversification in the internet economy is only the tip of the iceberg. 
Further investigation is likely to give better insights. 
For example, one natural constraint on the sponsored search auction comes from the fact that 
there is a limit on the number of slots, in particular for the popular keywords, which limits 
the number of advertisers that can be accommodated and it is likely that new market mechanisms as well as new 
for-profit agents will emerge to combat or to make profit from the opportunities created by this
capacity constraint, leading to a diversification in the market.
We are investigating this direction in a related work\cite{Cap-Med}.
Further, even in the case of for-profit mediators studied in the present work, 
several directions are left to explore. For example, do there exist mechanisms 
which are impervious to collusion via for-profit mediation and if they do 
how do they effect the revenue of the auctioneer? Furthermore, more sophisticated 
mediators such as where $M$-bidders need not be concecutive should also be investigated 
and in general if the mediator is sophisticated enough to exploit her best 
possible strategy, how does the modified profile look at equilibrium?
Furthermore, for-profit mediators for other auction formats should also be interesting to 
study.

\end{document}